\documentclass[aps,prl,superscriptaddress,twocolumn]{revtex4-2}
\usepackage[utf8]{inputenc}
\usepackage{graphicx,braket,amsmath,amssymb,nicefrac}

\begin{document}

\title{Experimental Measurement of Out-of-Time-Ordered Correlators at Finite Temperature}

\author{Alaina M. Green}
\affiliation{Joint Quantum Institute and Department of Physics, University of Maryland, College Park, MD 20742, USA}
\author{A. Elben}
	\affiliation{Institute for Quantum Information and Matter and Walter Burke Institute for
Theoretical Physics, California Institute of Technology, Pasadena, CA 91125, USA}
	\affiliation{Center for Quantum Physics, University of Innsbruck, Innsbruck A-6020, Austria}	
\affiliation{Institute for Quantum Optics and Quantum Information of the Austrian Academy of Sciences,  Innsbruck A-6020, Austria}
\author{C. Huerta Alderete}
\affiliation{Joint Quantum Institute and Department of Physics, University of Maryland, College Park, MD 20742, USA}
\author{Lata Kh Joshi}
\affiliation{Center for Quantum Physics, University of Innsbruck, Innsbruck A-6020, Austria}
\affiliation{Institute for Quantum Optics and Quantum Information of the Austrian Academy
of Sciences, Innsbruck A-6020, Austria}
\author{\mbox{Nhung H. Nguyen}}
\affiliation{Joint Quantum Institute and Department of Physics, University of Maryland, College Park, MD 20742, USA}
\author{Torsten V. Zache}
\affiliation{Center for Quantum Physics, University of Innsbruck, Innsbruck A-6020, Austria}
\affiliation{Institute for Quantum Optics and Quantum Information of the Austrian Academy
of Sciences, Innsbruck A-6020, Austria}
\author{Yingyue Zhu}
\affiliation{Joint Quantum Institute and Department of Physics, University of Maryland, College Park, MD 20742, USA}
\author{Bhuvanesh Sundar}
\affiliation{Institute for Quantum Optics and Quantum Information of the Austrian Academy
of Sciences, Innsbruck A-6020, Austria}
\affiliation{JILA, Department of Physics, University of Colorado, Boulder, CO 80309, USA}
\author{Norbert M. Linke}
\affiliation{Joint Quantum Institute and Department of Physics, University of Maryland, College Park, MD 20742, USA}

\date{\today}

\begin{abstract}

Out-of-time-ordered correlators (OTOCs) are a key observable in a wide range of interconnected fields including many-body physics, quantum information science, and quantum gravity. 
Measuring OTOCs using near-term quantum simulators will extend our ability to explore fundamental aspects of these fields and the subtle connections between them. Here, we demonstrate an experimental method to measure OTOCs at finite temperatures and use the method to study their temperature dependence. These measurements are performed on a digital quantum computer running a simulation of the transverse field Ising model. Our flexible method, based on the creation of a thermofield double state, can be extended to other models and enables us to probe the OTOC's temperature-dependent decay rate. Measuring this decay rate opens up the possibility of testing the fundamental temperature-dependent bounds on quantum information scrambling.

\end{abstract}

\maketitle

\section{Introduction}

A key piece in our understanding of quantum many-body dynamics is the scrambling of quantum information. 
Out-of-time-ordered correlators (OTOCs) are powerful tools to probe quantum information scrambling in quantum many-body systems. OTOCs are multipoint correlation functions evaluated between operators with the probe times appearing out of order~\cite{swi18}. 
In particular, the decay of the magnitude of the four-point OTOCs with time indicates the distribution of an initially local perturbation over the system's degrees of freedom, i.e. the scrambling of quantum information. 
In analogy to chaotic classical systems which are characterized by their sensitivity towards small perturbations, OTOCs were originally 
recognized in high-energy physics literature~\cite{mal16bound, she142, she141} as probes of quantum chaos. They are now routinely used as tools to study the dynamics of many-body quantum systems~\cite{fan17,che17,cho17,he17,boh17,sch21,hua17}, out-of-equilibrium fluctuations~\cite{cam17,yun17,yun18}, quantum phase transitions~\cite{lewis2020detecting, nie2020experimental, daug2019detection1, daug2020topologically, wei2019dynamical, sun2020out, wang2019probing, heyl2018detecting, shen2017out}, and to elucidate quantum information spreading in black holes through the AdS/CFT correspondence~\cite{mal99,she141,she142,mal16syk,mal16bound,mal17,mal18, review}. 

It is of fundamental importance to measure OTOCs at finite temperature to study the effects of temperature on information scrambling. The advancement of techniques for measuring thermal OTOCs will serve to experimentally verify the theoretical conjecture on the maximum rate of scrambling~\cite{mal16bound, murthy2019bounds}.


There have recently been several theoretical proposals~\cite{yao16,dre18,ver19,yos19,lan20,sun21} to measure the thermal OTOC, but none have been realized so far. Measuring OTOCs in experiment poses a few stringent requirements, such as the realization of coherent backward time evolution - which is typically required due to the out-of-time ordering of the operators in its definition - and preparation of systems at controllable finite temperature. Previous experiments have measured OTOCs 
in pure initial states or at infinite temperature in several platforms including NMR~\cite{li17,wei18}, superconducting circuits~\cite{mi21,bra21}, trapped ions~\cite{gar17ion,lan19,jos20}, and cold atoms~\cite{pegahan2021energy}. While all these experiments relied on the ability to realize backward time evolution, OTOCs have also been measured at infinite temperature using randomized measurements without requiring backward time evolution~\cite{jos20, nie19}.


In this paper, we demonstrate the experimental measurement of thermal OTOCs using a digital quantum computer based on trapped ions. Our platform allows individual control of each qubit, and implements a universal gate set that can simulate arbitrary quantum dynamics as Trotterized Hamiltonian evolution~\cite{zhu21}. We prepare a thermal state on a subsystem of a larger pure state, constituting a thermofield double (TFD) state~\cite{tak96}, using variational quantum circuits as previously demonstrated in Refs.~\cite{wu19, zhu19}. These capabilities allow us to measure thermal OTOCs in a transverse-field Ising chain of three qubits, based on earlier theoretical proposals~\cite{lan20,sun21}. We measure information scrambling by tracking the decay of the magnitude of the thermal OTOC with time. As a key result, we demonstrate that temperature plays a significant role in scrambling, and that information scrambles faster at higher temperatures. 

\begin{figure*}[t]
	\includegraphics[width=1.97\columnwidth]{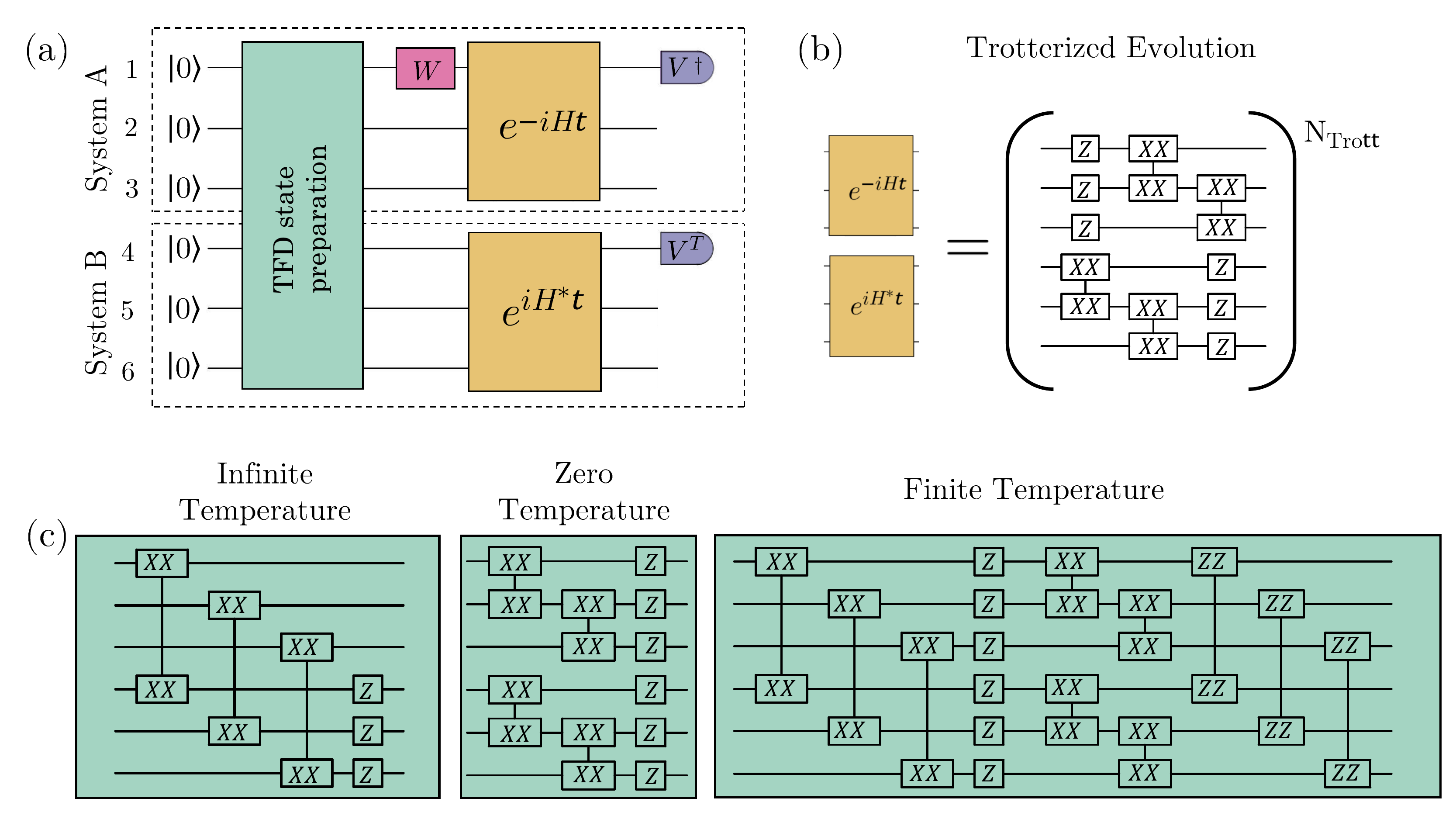}
\caption{An overview of the circuit executed on the trapped-ion quantum computer for OTOC measurement at various temperatures in the TFIM. Two copies of the three-spin system are represented by six qubits as shown in (a). From the $\ket{0}^{\otimes N}$ state, the TFD state is prepared using the circuits appropriate for the chosen temperature shown in (c). For the finite temperature TFD state, the set of angles assigned to the gates shown determine the temperature. The perturbing operator $\hat{W}$ imparts local information onto one copy system and each copy is evolved with the appropriate Hamiltonian digitized as shown in (b). Finally, correlations between the two system copies, $\langle \hat{V}^\dagger \otimes \hat{V}^T \rangle$, are measured to determine the OTOC. Here  $XX = e^{-i\frac{\theta}{2}\hat{\sigma}^x_{i}\hat{\sigma}^x_{j}}$, $ZZ = e^{-i\frac{\theta}{2}\hat{\sigma}^z_{i}\hat{\sigma}^z_{j}}$, and $Z=e^{-i \frac{\theta}{2}\hat{\sigma}^z_i}$, where $i$ and $j$ are qubit indices and $\theta$ is the rotation angle. The values of $\theta$ are given in~\cite{suppmat}.}
	\label{fig:Circuit}
\end{figure*}

\section{Methods and Model}

Our goal is to measure the following form of the thermal OTOC at inverse temperature $\beta=1/T$ for a Hamiltonian $\hat{H}$:
\begin{multline}\label{eq:Odefinition}
    O=\frac{1}{Z} \text{Tr}\left(e^{-\beta \hat{H}/2} \hat{W}^\dagger \hat{V}^{\dagger}(t) \hat{W} e^{-\beta \hat{H}/2} \hat{V}(t) \right),
\end{multline}
\noindent where $\hat{V}(t)=e^{i\hat{H}t}\hat{V}e^{-i\hat{H}t}$ is Hermitian and $\hat{W}$ is a local unitary operator. The normalization factor is given by $Z=\text{Tr}[e^{-\beta \hat{H}}]$. We take $\hbar=k_B=1$.

The experimental sequence for measuring $O$, based on previous theoretical proposals~\cite{sun21, lan20}
, is summarized in Fig.~\ref{fig:Circuit}(a). Central to our method is the creation of a TFD state spanning two system copies labeled $A$ and $B$, each of $N$ qubits, and taking the form
\begin{equation}
    \ket{\psi_{\mathrm{TFD}}}=\frac{1}{\sqrt{Z}}\sum_{n}e^{-\beta E_{n}/2}\ket{n}_A\ket{n^*}_B,
\end{equation}
\noindent where $n$ labels the eigenstates of $\hat{H}$ in each copy with energies $E_n$, and the star indicates the complex conjugate state.  Although $\ket{\psi_{\mathrm{TFD}}}$ is a pure state, expectation values taken with respect to the TFD state give thermally averaged results for each system copy. Denoting the Hamiltonian for each system copy as $\hat{H}_A$ and $\hat{H}_B$, and the initial perturbation as $\hat{W}$, we first apply $\hat{W}_A$ on copy $A$ in the initial state, and then evolve the two-copy system with $\hat{H}_A - \hat{H}_B^*$ as shown in Fig.~\ref{fig:Circuit}(a). The state just after time evolution is given by:
\begin{equation}
   \ket{\psi\left(t\right)} = e^{-i \left(\hat{H}_{A}-\hat{H}_{B}^*\right) t} \hat{W}_{A} \ket{\psi_{\mathrm{TFD}}}.
\end{equation}
Operators $\hat{V}_A$ and $\hat{V}_B$ are mirrored between the two copies and constitute the local measurements giving access to the OTOC in the form of their correlator:
\begin{multline}\label{eq:otocderivation}
    \bra{\psi(t)} \hat{V}_A^\dagger \otimes \hat{V}_B^T \ket{\psi(t)}= \\ \frac{1}{Z}\sum_{n,m} \bra{m}_A \bra{m^*}_B e^{-\beta E_{m}/2} \hat{W}_A^{\dagger} e^{i \left(\hat{H}_{A}-\hat{H}_{B}^*\right) t} \hat{V}_A^\dagger \otimes \\ \hat{V}_B^T e^{-i \left(\hat{H}_{A}-\hat{H}_{B}^*\right) t} e^{-\beta E_{n}/2} \hat{W}_A \ket{n}_A\ket{n^*}_B.
\end{multline}
\noindent We split the expectation value in the joint Hilbert space of $A$ and $B$ in Eq.~\eqref{eq:otocderivation} into a product of two matrix elements separately in the $A$ and $B$ Hilbert spaces, and hereafter drop the labels. We obtain

\begin{align}
    \label{eq:productterms}
    \bra{\psi(t)} \hat{V}^\dagger \otimes \hat{V}^T \ket{\psi(t)}=  \frac{1}{Z}\sum_{n,m} \braket{m^* \vert e^{-i \hat{H}^* t} \hat{V}^T e^{i \hat{H}^* t} \vert n^*}\nonumber\\
    \times \braket{m \vert e^{-\beta E_{m}/2} \hat{W}^{\dagger} e^{i \hat{H} t} \hat{V}^\dagger e^{-i \hat{H} t} e^{-\beta E_{n}/2} \hat{W} \vert n}.
\end{align}

\noindent Subsequently, we utilize the relations $e^{-\beta E_n/2}\ket{n} = e^{-\beta \hat{H}/2}\ket{n}$ and $\bra{m^*} e^{-i\hat{H}^*t} \hat{V}^T e^{i\hat{H}^*t} \ket{n^*} = \bra{n}\hat{V}(t)\ket{m}$ to obtain:
\begin{multline}
    \label{eq:OSum}
    \bra{\psi(t)} \hat{V}^\dagger \otimes \hat{V}^T \ket{\psi(t)}= \\ \frac{1}{Z}\sum_{n,m} \bra{m} e^{-\beta \hat{H}/2} \hat{W}^{\dagger} \hat{V}^\dagger(t) \hat{W} e^{-\beta \hat{H}/2} \ket{n} \bra{n}\hat{V}(t)\ket{m}.
\end{multline}

\noindent Resolving the identity operator in Eq.~\eqref{eq:OSum}~\footnote{After dropping the unnecessary labels $A$ and $B$, it is irrelevant that $\ket{n}_A$ and $\ket{n^*}_B$ came from different subsystems. Therefore, the term $\sum_n \ket{n}\bra{n}$ is equal to the identity operator.}, $\sum_n \ket{n}\bra{n} = \hat{1}$, we find that the right-hand side of Eq.~\eqref{eq:OSum} reduces to $O$ in Eq.~\eqref{eq:Odefinition}.




To demonstrate our method, we choose to study OTOCs in the transverse-field Ising model (TFIM), which can be very efficiently simulated on a trapped-ion quantum computer (TIQC). The Hamiltonian for this model is:
\begin{equation}
    \hat{H}=J\sum^{N-1}_{i=1} \hat{\sigma}^x_{i}\hat{\sigma}^x_{i+1}+g\sum^{N}_{i=1}\hat{\sigma}^z_i
    \label{eq:TFIM},
\end{equation}
where $J$ gives the spin-spin coupling strength, $g$ determines the strength of the interaction with the field, and $\hat{\sigma}^{\alpha}_i~,\alpha\in(x,y,z)$ are Pauli spin operators on the $i$th site. Note that $\hat{H}^*=\hat{H}$. Time evolution is performed through Trotterization, in which the full evolution is approximated by executing one Trotter step sequentially $N_{\mathrm{Trott}}$ times. The digital circuit for each Trotter step is represented in brackets in Fig.~\ref{fig:Circuit}(b). It is constructed from single-qubit rotations and two-qubit entangling gates. The parameters for these gates, i.e. their angles of rotation on the single or two-qubit Bloch sphere, determine the evolution time spanned by a single Trotter step.


In order to prepare the TFD state, we follow the variational method in Refs.~\cite{wu19,zhu19} with classically optimized parameters. This method digitally approximates the true thermal state. We use classical simulation of the quantum circuit to quantify this approximation and find that, in the absence of physical errors, this method would allow us to create TFD states with fidelity greater than 97\% for all temperatures considered in this work. However, physical errors limit the fidelity of our experimentally prepared state to less than 97\%. Note that in the case of infinite temperature, the theoretical preparation circuit requires no approximation. For the finite temperature preparation circuits, the exact parameters used for the gates determine the temperature of the OTOC measurement.

We execute these circuits on a TIQC, which has previously been described in Ref.~\cite{deb16}. The TIQC is built upon a trapped linear chain of $^{171}\mathrm{Yb}^{+}$ ions with the qubit encoded in two different hyperfine ground states. All qubits are initialized in the $\ket{0}$ state via optical pumping~\cite{olm07}. Coherent operations are implemented using a pair of Raman beams derived from a mode-locked laser at 355 nm. The beams are counterpropagating and one of them is split into an array of tightly focused beams, each addressing a single ion in the chain. Single-qubit gates are compiled into pulses of appropriate phase, amplitude and duration to effect a resonant Rabi rotation while two-qubit gates are compiled into amplitude and frequency modulated pulses which effectively entangle the qubits' spin degrees of freedom through transient entanglement of both spins with the shared motion of the ions in the Paul trap~\cite{mol99,blu21}. The typical fidelities of single- and two-qubit gates are 99.5\% and 98.5\%, respectively. Measurements are performed in the computational basis with state-dependent fluorescence detection~\cite{olm07}. The entire experiment is repeated several thousand times to detect the average state population, which is corrected for errors of about $1\%$ arising from imperfect state preparation and measurement that are independently characterized.

\section{Results}

\begin{figure}
	\includegraphics[width=0.97\columnwidth]{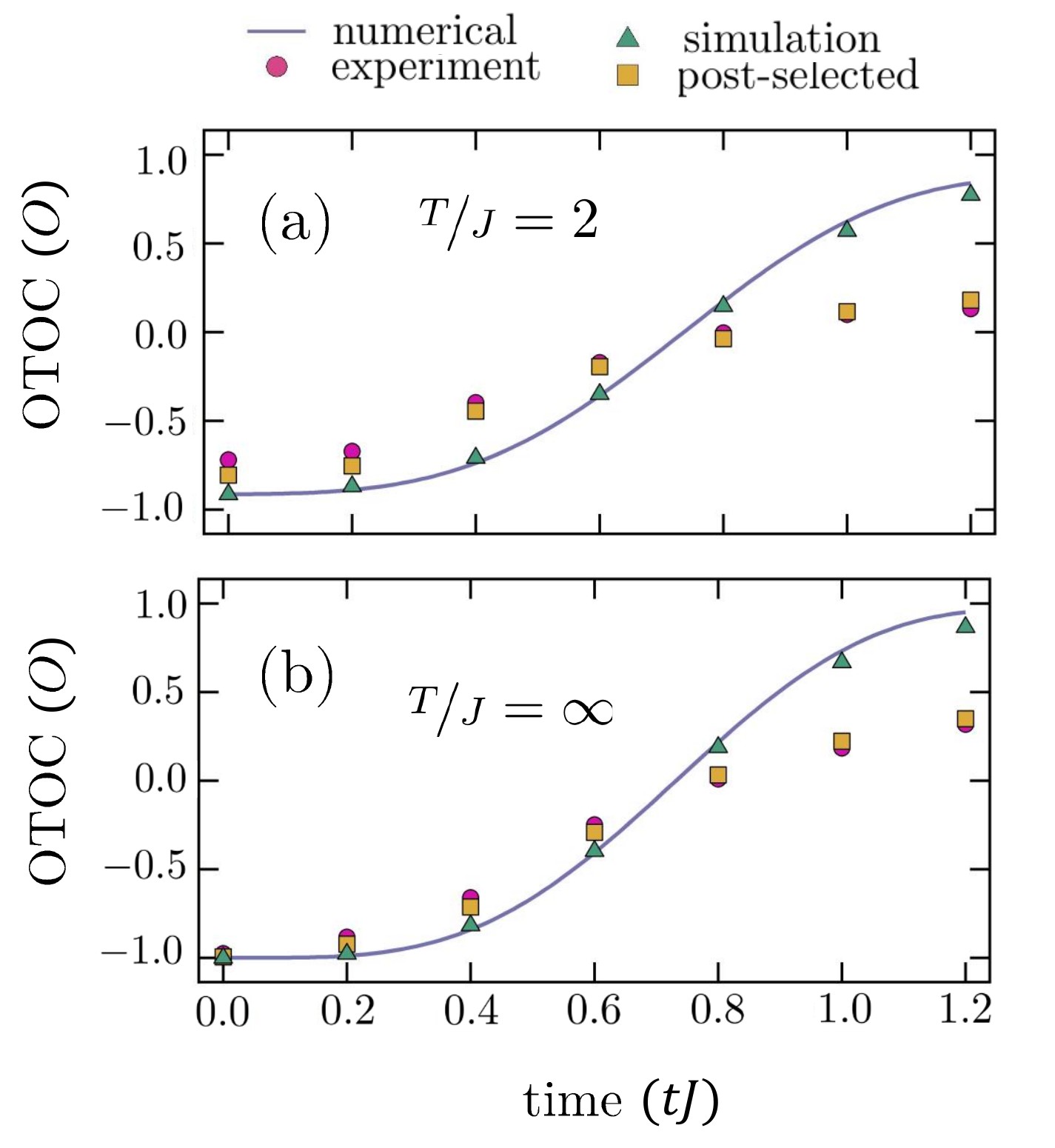}
    \caption{The evolution of the OTOC is measured as summarized in Fig.~\ref{fig:Circuit}. Here we have prepared the TFD state for $T/J=2$ in (a) $T/J=\infty$ in (b) and chosen $J=g$ (see Eq.~\eqref{eq:TFIM}). In both cases we use $\langle \hat{V}^\dagger \otimes \hat{V}^T \rangle=\langle \hat{\sigma}_1^{z}\otimes \hat{\sigma}_4^{z} \rangle$ for our measurement and have used $\hat{W}=\hat{\sigma}^x_1$ to impart local information. Pink circles show the experimental result, yellow squares the experimental result after postselection (see text), green triangles the expected results in an error-free circuit, and the purple line corresponds to numerical simulation of the model. For the experimental values, the statistical error bars are smaller than the symbols. While the two OTOCs at both temperatures exhibit qualitatively similar decay, close inspection shows a significant difference in the rate of decay (see Fig.~\ref{fig:TempDependence}).}
	\label{fig:SampleEvolution}
\end{figure}


We measure the OTOCs in the TFIM at different times, set by our Trotter steps, by executing the circuits in Fig.~\ref{fig:Circuit}. Fig.~\ref{fig:SampleEvolution} shows time evolution of the OTOC for two different temperatures. In this case we choose $\hat{W}=\hat{\sigma}_1^x$ and we measure the correlator $\langle \hat{V}^{\dagger}\otimes \hat{V}^{T} \rangle=\langle \hat{\sigma}_1^{z}\otimes \hat{\sigma}_4^{z} \rangle$, although any corresponding pair between the two copies of the system could be used. We compare the experimentally measured results to those of the expected results which can be calculated exactly in this small demonstration. The observed decay of the OTOC's magnitude from its initial temperature-dependent value agrees well with the expected result. The deviation between the measured results (pink circles) and the exact evolution (purple line) can be attributed to two general errors: those arising from algorithmic approximations and those arising from physical errors on the apparatus. Algorithmic errors comprise both imperfect variational preparation of the TFD state and Trotter error, and can be read in Fig.~\ref{fig:SampleEvolution} as the difference between the purple line and the green triangles. As can be seen in the Fig.~\ref{fig:SampleEvolution}, algorithmic error is small compared to the physical error which is likely dominated by imperfect calibration of the gates or residual entanglement of the qubit states with the ion motion. 

A small portion of the physical error can be eliminated through postselection whereby we take advantage of the symmetry of our model to discard results which could only arise through physical error. The symmetry of the TFIM dictates that its eigenstates satisfy $\prod^{2N}_{i=1}\hat{\sigma}_i^z \ket{n}\ket{n^*}=\ket{n}\ket{n^*}$, and we therefore have 
$\langle \psi_{\rm TFD}\vert \prod^{2N}_{i=1}\hat{\sigma}_i^z \vert\psi_{\rm TFD}\rangle=1$ and $\langle \psi_{\rm TFD}\vert \hat{\sigma}^x_1 \left(\prod^{2N}_{i=1}\hat{\sigma}_i^z\right) \hat{\sigma}^x_1 \vert\psi_{\rm TFD}\rangle=-1$. 
Noting that the initial state before time evolution is $\ket{\psi(0)} = \hat{\sigma}^x_1 \ket{\psi_{\rm TFD}}$, and because $\prod^{2N}_{i=1}\hat{\sigma}^{z}_i$ commutes with $\hat{H}_A-\hat{H}_{B}^*$, any experimental results for which $\langle \prod^{2N}_{i=1}\hat{\sigma}_i^z\rangle \ne-1$ may be discarded. For the entirety of the results shown in Figs.~\ref{fig:SampleEvolution}(a) and (b) respectively, an average of 40\% and 20\% of the measurements are discarded in postselection. 
While the value of $\langle \hat{V}^{\dagger}\otimes \hat{V}^{T} \rangle$ grows from an initial value near $-1$, i.e. its magnitude decays, as expected by the numerical simulation, the decay is damped through physical errors. Separating further physical errors from genuine decay of the OTOC will be treated heuristically below.


As our main result, we show the dependence of the OTOC decay rate with temperature. In order to quantify this, we approximate the derivative, $O'(t)$, by the finite difference $\lambda$ at the point where the decay of the OTOC is expected to be largest. In particular, we extract $\lambda \equiv \frac{O\left(0.8/J\right)-O\left(0.4/J\right)}{0.4/J}$. A fair comparison of the OTOC magnitudes' decay at different temperatures requires a similar amount of physical error in all the circuits. To accomplish this, we artificially add gates to the shorter depth circuits such that the number of gates are the same for a given time. In particular, we add to state preparation extra gates which do not affect the resulting state except through physical error. 

\begin{figure}
	\includegraphics[width=0.97\columnwidth]{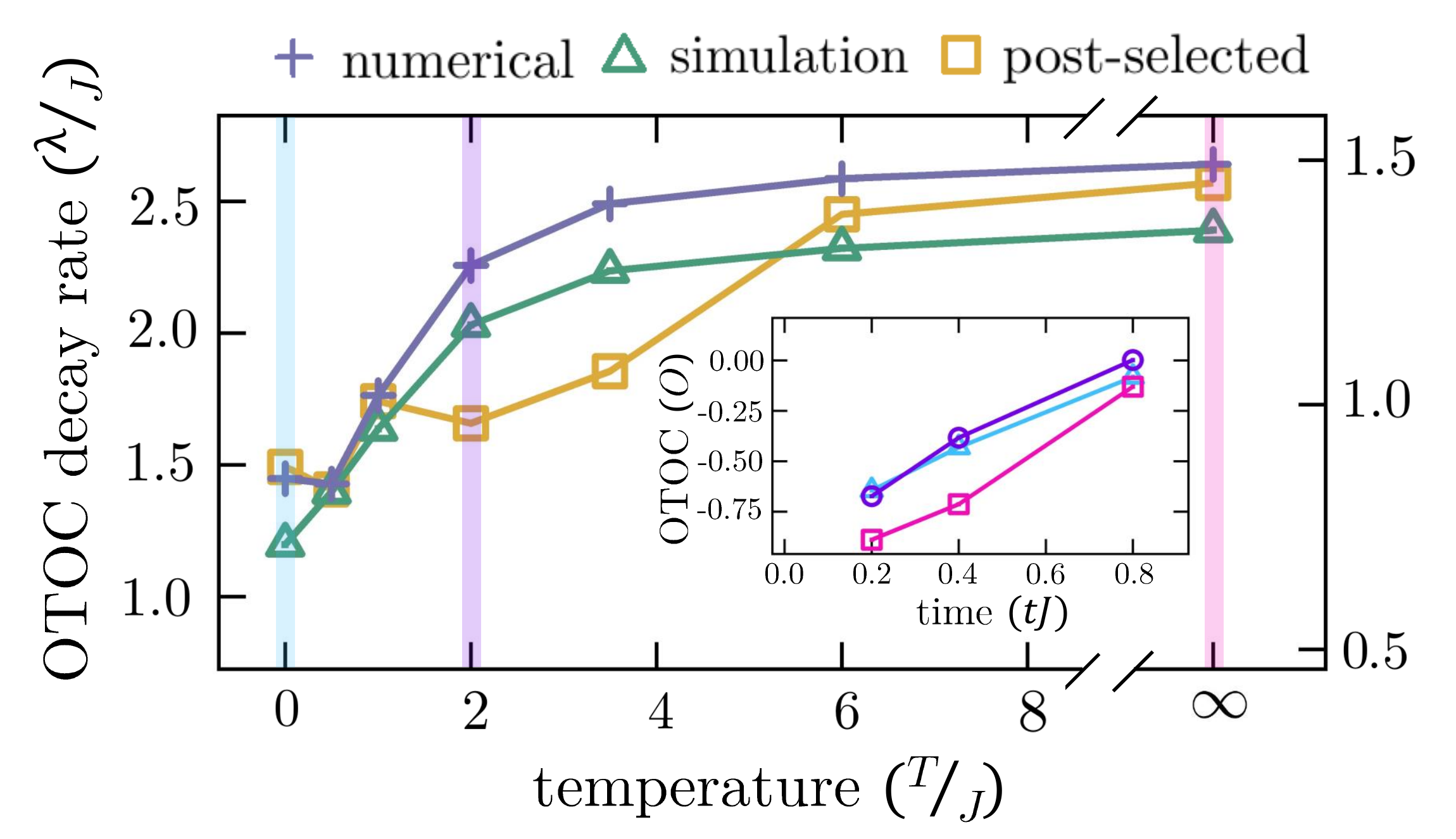}
    \caption{Temperature-dependence of the OTOC decay rate $O'$, approximated by the finite difference $\lambda$ (see main text). On the right axis, yellow squares show the experimental results after postselection (see text). On the left axis, green triangles show simulation of the circuit on an error-free quantum computer and the purple plus signs show the slope for the numerical simulation. (inset) Sample evolution of $\langle \hat{V}^{\dagger}\otimes \hat{V}^{T} \rangle$ at zero temperature (light blue triangles), for $T/J=2$ (purple circles) and infinite temperature (pink squares), showing decay of the OTOC magnitude. The Trotter step size is not uniform but proceeds in sizes $\{0.2,0.2,0.4\}$ $tJ$. For the experimental data the statistical error bars are smaller than the symbols.}
	\label{fig:TempDependence}
\end{figure}

Fig.~\ref{fig:TempDependence} shows the OTOC decay rate versus temperature, with the inset showing the magnitude's decay versus time at a few sample temperatures. As expected from the sample results in Fig.~\ref{fig:SampleEvolution}, the experimentally measured decay rate is lower than the theoretically expected result. In order to better compare the change in decay rate with temperature to theoretical expectations, we choose to plot the experimental decay rate on a different axis than the theoretical ones. The OTOC decay rate is experimentally observed to increase with temperature with a trend similar to that seen in the numerical results.

\section{Outlook}


Our measurement of the temperature dependence of OTOC decay represents a new tool for digital quantum simulation, providing a measurable quantity which can be used to study the temperature dependence of quantum information scrambling. In particular, experimental access to thermal OTOCs provides a tantalizing route to using quantum simulators to test fast scrambling and bounds on the rate of information scrambling~\cite{mal16bound, murthy2019bounds}. 
This includes testing the exponential decay of OTOCs in fast-scrambling quantum models~\cite{belyansky2020minimal, li2020fast, yin2020bound}, and saturation of the bound in models that are analogous to black holes via holographic duality~\cite{mal99, mal16syk}.
Additional building blocks for realizing such a model have already been considered in Refs.~\cite{gar17ads,mar19,su21}. Measuring thermal OTOCs also provides a route to implement and benchmark ideas for simulating many-body teleportation and traversable wormholes in the lab~\cite{sch21, bro21, nezami2021quantum}.

\section{Acknowledgements}
	A.E.\ acknowledges funding by the German National Academy of Sciences Leopoldina under the grant number LPDS 2021-02 and  by the Walter Burke
Institute for Theoretical Physics, Caltech. L.K.J. acknowledges the European Union's Horizon 2020 research and innovation programme under Grant Agreement  No.\ 731473 (QuantERA via QT-FLAG) and the Austrian Science Foundation (FWF, P 32597 N). T.V.Z.'s work is supported by the Simons Collaboration on Ultra-Quantum Matter, which is a grant from the Simons Foundation (651440, P.Z.). This work received support from the National Science Foundation through the Quantum Leap Challenge Institute for Robust Quantum Simulation (OMA-2120757) and the Physics Frontier Center (PHY-1430094) at the Joint Quantum Institute (JQI). A.M.G. is supported by a JQI Postdoctoral Fellowship. N.M.L. acknowledges funding by the Maryland-Army-Research-Lab Quantum Partnership (W911NF1920181), the Department of Energy, Office of Science, Office of Nuclear Physics (DE-SC0021143), and the Office of Naval Research (N00014-20-1-2695). We thank Ana Maria Rey and Murray Holland for a careful reading of the manuscript.

\end{document}